\documentstyle[12pt]{article}
\def\be{\begin{equation}}
\def\ee{\end{equation}}
\newtheorem{Theorem}{Theorem}
\begin{document}
\rightline{\tt ITFA-25/92}
\vskip 3cm
\centerline{\LARGE Finite W symmetry in finite dimensional}
\centerline{\LARGE integrable systems \normalsize\footnote{talk given at the
'workshop on low dimensional topology and physics'}}
\vskip 1cm
\centerline{\large T.Tjin \normalsize\footnote{email: tjin@phys.uva.nl}}
\vskip 0.5cm
\centerline{Institute for Theoretical Physics, University of Amsterdam}
\vskip2pt
\centerline{Valckenierstraat 65, 1018 XE Amsterdam}
\vskip2pt
\centerline{The Netherlands}
\vskip2pt
\centerline{October 1992}
\vskip 1cm

\begin{abstract}
By generalizing the Drinfeld-Sokolov reduction a large class of
$W$ algebras can be constructed. We introduce 'finite' versions of
these algebras by Poisson reducing Kirillov Poisson structures on
simple Lie algebras.
A closed and coordinate free formula for the reduced Poisson
structure is given.
These finitely generated algebras play the same
role in the theory of $W$ algebras as the simple Lie algebras
in the theory of Kac-Moody algebras and will therefore presumably
play an important role in the representation theory of $W$ algebras.
We give an example leading to
a quadratic $sl_2$ algebra. The finite dimensional unitary representations
of this algebra are discussed and it is shown that they have
Fock realizations. It is also shown that finite dimensional
generalized Toda theories are reductions of a system describing
a free particle on a group manifold. These
finite  Toda systems have the non-linear finite $W$ symmetry discussed
above.
\end{abstract}
\pagebreak

\section*{Introduction}
Nonlinear extensions of the Virasoro algebra, generally known
as $W$ algebras, have turned up in various areas of
mathematical physics (see \cite{BS} for a
recent review). Unfortunately not too much is known about them, their
interpretation, classification and representation
theory is still far from complete.
In this paper we discuss a very simple class of nonlinear algebras,
which are basically 'finite' $W$ algebras and which may serve
as instructive playground for the infinite case. However as we
shall see they do have some interest of their own.

In \cite{D1} $W_n$ algebras, were shown to arize as
Dirac bracket algebras
on submanifolds of Kac-Moody algebras. This gave a clear understanding
of how nonlinear $W$ algebras can arize as Poisson reductions
of linear current algebras. However, $W_n$ algebras are certainly not
the only $W$ algebras known in the literature, which leads one
to ask whether others could be constructed in a similar way.
The answer to this question turned out to be yes as it was
shown in \cite{BTV} that there are as many different Poisson reductions
of an $sl_n$ current algebra leading to $W$ algebras as there are
partitions of the number $n$. The reduction point of view
has recently been investigated by a great number of people (for
example
\cite{D1,BTV,D2,D3,D4,D5,D6,D7,BO,FF,So1,So2}).
The classical covariant
$W$ gravity theories for all the algebras which can be constructed as
reductions of Kac-Moody algebras
as well as their moduli spaces have been constructed in \cite{BG}.

Having the picture of $W$ algebras as reductions of
Kac-Moody algebras in mind one can ask
oneself the question whether this is special for infinitely
many dimensions, i.e. can one formulate a similar theory
for finite dimensional Lie algebras?  The answer to this
question is affirmative as was shown in \cite{T}. This lead to so called
'finite $W$ algebras'. It turns out that these are a very effective
toy model for ordinary $W$ algebras but apart from that they
are of some interest themselves. For
example they are what one could consider to be the finite algebras
underlying the $W$ algebras, just as the simple Lie algebras underly
the Kac-Moody algebras. This means that they will play an important
role in the representation theory of ordinary $W$ algebras since
the subspace of singular vectors of a $W$ algebra will carry a
representation of the underlying finite $W$ algebra.

Another area
where finite $W$ algebras may play an important role is finite
dimensional generalized Toda theory. These were originally
introduced as dimensional reductions
of self dual Yang-Mills theories in \cite{BV}. The general solution
space of these models was however not constructed (only for some
special examples). In this paper we will show that
these generalized finite Toda systems are reductions of a system
describing a free particle moving on a group manifold and that
they have the finite $W$ symmetry discussed above. The
general solution of the equations of motion of such a system
is easily found and has the form of a transform by the
symmetry group of a certain reference solution. This reference
solution can be reduced to give a nontrivial solution of the
generalized Toda theory and one would expect the orbits of
finite $W$ transformations to provide the other solutions.

In this paper we shall give a brief account of  the ideas
described above deferring details to our forthcoming paper \cite{BGT}.

\section{Generalized Drinfeld-Sokolov reductions}
In this section we briefly discuss the results of \cite{BTV}
in which the Drinfeld-Sokolov reductions were generalized.

Let $g$ be simple Lie algebra, $\{t_a\}$ a basis of $g$ and $f_{ab}^c$
the structure constants of $g$ in this basis. Consider the Poisson
algebra
\be
\{J^a(z),J^b(w)\}=f^{ab}_cJ^c(w)\delta (z-w) -kg^{ab}\delta '(z-w)
\label{KM}
\ee
where $g^{ab}$ is the inverse matrix of $g_{ab}=\mbox{Tr}(t_at_b)$.
This Poisson structure is actualy nothing but the Kirillov Poisson
structure on the affine Kac-Moody algebra over $g$.
{}From now on we fix $g=sl_n$.

Let $i: sl_2 \hookrightarrow sl_n$ be an embedding of $sl_2$ into
$sl_n$. Under the adjoint action of the embedded $sl_2$ algebra
the algebra $sl_n$ branches into a direct sum of $p$ irreducible
$sl_2$ multiplets. Let $\{t_{k,m}\}_{m=-j_k}^{j_k}$ be a basis
of the $k^{th}$ multiplet where $j_k$ is the highest weight of
this multiplet. The numbering is chosen such that $t_{1,\pm 1}=
t_{\pm}$ and $t_{1,0}=t_3$ where $\{t_3,t_{\pm}\}$ are the
Cartan, step up and step down elements of $i(sl_2)$. An arbitrary
map $J:S^1 \rightarrow sl_n$ can then be written as
\be
J(z)=\sum_{k=1}^p \sum_{m=-j_k}^{j_k}J^{k,m}t_{k,m} \label{un}
\ee
Impose now the constraints $\phi^{1,1}(z)\equiv J^{1,1}(z)-1=0$
and $\phi^{k,m}(z)=J^{k,m}=0$ for $m>0, \;k \neq 1$. The constraints
$\{\phi^{k,m}(z)\}_{m\leq 1}$ are first class which means they generate
gauge invariance. This gauge invariance can be completely fixed by
gauging away the fields $\{J^{k,m}(z)\}_{m >-j_k}$. After constraining
and gauge fixing the currents look like
\be
J_{fix}(z)=\sum_{k=1}^p J^{k,-j_k}t_{k,-j_k}+t_+    \label{fix}
\ee
The Poisson bracket (\ref{KM}) on the set of 'currents' of the
form (\ref{un}) induces a Poisson bracket (which is in fact a Dirac
bracket as first realized in
\cite{D1}) on the set of gauge fixed currents (\ref{fix}). The
algebra generated by the fields $\{J^{k,-j_k}(z)\}$ and equipped
with the Dirac bracket is then a $W$ algebra with conformal
weights $\{\Delta_k=j_k+1\}$.

The ordinary Drinfeld-Sokolov reductions which yield the Zamolodchikov
$W_n$ algebras correspond to the case where one takes the principal
embedding of $sl_2$ into $sl_n$. The algebra $W_3^{(2)}$ introduced
first by Polyakov and Bershadsky corresponds to the only non-principal
$sl_2$ embedding into $sl_3$. In general however there are as
many inequivalent $sl_2$ embeddings into $sl_n$
as there are partitions of the number $n$. This gives a large
number of possibilities.

\section{Finite W algebras}
In this section we introduce finite $W$ algebras \cite{T} by reducing
finite dimensional simple Lie algebras instead of KM algebras

The starting point is again the Kirillov Poisson structure
on a Lie algebra (actually it is on its dual but since simple
Lie algebras carry a nondegenerate bilinear form we identify the
Lie algebra with its dual). The coordinate free expression of
this Poisson bracket is
\be
\{F,G\}(x) = \left( x, [\mbox{grad}_xF,\mbox{grad}_xG] \right)
\ee
where $(.,.)$ is the Cartan-Killing form, $F,G$ are
smooth functions on $g$ and  $grad_xF$ is
defined by
\be
\frac{d}{d \epsilon }F(x+\epsilon x')|_{\epsilon =0}=
\left( x',\mbox{grad}_xF \right) \;\;\;\; \mbox{for all }x' \in g
\ee
Using again the basis $\{t_a\}$ an arbitrary element of $g$ can
be written as $x=J^a(x)t_a$ where $J^a$ is a smooth function on $g$.
In terms of these coordinate functions the Kirillov bracket reads
\be
\{J^a,J^b\}=f^{ab}_cJ^c
\ee
(compare to eq.(\ref{KM}).

One can go now through the whole procedure again, i.e. choose an
$sl_2$ embedding, impose the constraints and gauge fix.
Define the set of gauge fixed elements (which is a submanifold of $g$)
by
\be
g_{fix}=\{t_+ +\sum_{k=1}^p y^{k,-j_k}t_{k,-j_k}\mid y^{k,-j_k} \in
{\bf C,R}\}
\ee
Again the Kirillov Poisson bracket on $g$ induces a Poisson bracket
on $g_{fix}$. In order to describe this bracket introduce the
map
\be
L:g \longrightarrow g
\ee
which, on $\mbox{Im}(ad_{t_+})$ is the inverse of the map
$ad_{t_+}:\mbox{Im}(ad_{t_-}) \rightarrow \mbox{Im}(ad_{t_+})$ and
on the complement of $\mbox{Im}(ad_{t_+})$ is the zero map. It is
shown in \cite{BGT} that
for $Q_1$ and $Q_2$ smooth functions on $g_{fix}$ and $y\equiv t_++w \in
g_{fix}$ we have
\be
\{Q_1,Q_2\}(y)=\left( y,[\mbox{grad}_yQ_1,\frac{1}{1+L \circ ad_w}
\mbox{grad}_yQ_2] \right)    \label{red}
\ee
where $grad_yQ \in ker(ad_{t_+})$ is (uniquely) defined by
\be
\frac{d}{d \epsilon}Q(y+\epsilon y')|_{\epsilon =0}=
\left( y',\mbox{grad}_yQ \right)
\ee
for all $y' \in Ker(ad_{t_-})$. (This uniquely defines
$grad_yQ$ because $Ker(ad_{t_-})$ and $Ker(ad_{t_+})$ are nondegenerately
paired by the Cartan-Killing form.)

Let us consider an example. The finite versions of $W_n$,
corresponding to the principal $sl_2$ embeddings, give
abelian Poisson algebras and are therefore not very interesting.
The simplest nontrivial case is associated to the
nonprincipal $sl_2$ embedding of $sl_2$ into $sl_3$.
Under the adjoint action of this embedding $sl_3$ decomposes into
a direct sum of a triplet, two doublets and a singlet
(i.e. $k=1, \ldots, 4$ and $j_1=1,\; j_2=j_3=\frac{1}{2},\; j_4=0$).
The reduced
algebra will therefore have 4 generators
$J^{1,-1},J^{2,-1/2},J^{3,-1/2}$ and $J^{4,0}$
(or equivalently $g_{fix}$
is 4 dimensional).
The Poisson brackets (\ref{red})
in terms of $c=-\frac{4}{3}(J^{1,-1}+3(J^{4,0})^2)$,
$e=\sqrt{\frac{4}{3}}\;J^{2,1/2}$, $f=\sqrt{\frac{4}{3}}\;J^{3,-1/2}$
and $h=-4J^{4,0}$ read in this case
\begin{eqnarray}
\{h,e\} & = & 2e \nonumber \\
\{h,f\} & = & -2f \nonumber \\
\{e,f\} & = & h^2+c \label{alg}
\end{eqnarray}
and $c$ Poisson commutes with everything. This algebra is obviously
a non-linear and centrally extended
version of $sl_2$ and was first constructed in
\cite{Rocek} as a solution of the Jacobi identities. We summarize
the (real) representation theory of the commutator version
of this algebra in the following theorem \cite{T}.
\begin{Theorem}
Let $p$ be a positive integer and $x$ a real number.
\begin{enumerate}
\item For every pair $(p,x)$ the algebra (\ref{alg}) has a unique
highest weight representation $W(p;x)$ of dimension $p$ with
highest weight $j(p;x)=p+x-1$ and central value $c(p;x)=\frac{1}{3}
(1-p^2)-x^2$.
\item  Let $k \in \{1, \ldots , p-1\}$ then $W(p;\frac{2}{3}k-\frac{1}{3}
p)$ is reducible and its invariant subspace is isomorphic as
a representation to $W(p-k;-\frac{1}{3}(k+p))$.
\item The representation $W(p;x)$ is unitary iff $x>\frac{1}{3}p-
\frac{2}{3}$
\end{enumerate}
\end{Theorem}

It is well known that it is possible to realize the finite dimensional
irreducible representations of any simple Lie algebra on a Fock
space. Consider for example the realization on ${\bf C}[z]$ of the
algebra $sl_2$
\begin{eqnarray}
\sigma_{\Lambda}(t_+) & = & \frac{d}{dz} \nonumber \\
\sigma_{\Lambda}(t_-) & = & (\Lambda,\alpha )z-z^2\frac{d}{dz} \nonumber \\
\sigma_{\Lambda}(t_0) & = & (\Lambda,\alpha )-2z\frac{d}{dz}
\end{eqnarray}
where $\Lambda$ is a weight
and $\alpha$ is the positive root of $sl_2$. For $\Lambda$ a
principal dominant weight the representation $\sigma_{\Lambda}$
is reducible and  in fact the subspace
\be
V=\{P(z) \in {\bf C}[z] \mid
(\frac{d}{dz})^{(\Lambda,\alpha )+1}P(z)=0\}
\ee
is isomorphic to the $(\Lambda, \alpha )
+1$ dimensional irreducible representation of $sl_2$
(what we have presented here is the first term of a Fock resolution
of the $(\Lambda,\alpha )+1$ dimensional irrep of $sl_2$).

Similar realizations exist for the representations $W(p;x)$ of
the algebra (\ref{alg}). Define the representation $\hat{\sigma}_{
\Lambda}$ by
\begin{eqnarray}
\hat{\sigma}_{\Lambda}(h) & = & (\Lambda,\alpha_1-\alpha_2 )+
\frac{1}{3} -2z\frac{d}{dz} \nonumber \\
\hat{\sigma}_{\Lambda}(e) & = & -3(\Lambda , \alpha_2 )
\frac{d}{dz}-2z\frac{d^2}{dz^2} \nonumber \\
\hat{\sigma}_{\Lambda}(f) & = & (\Lambda , \alpha_1 ) z -
\frac{2}{3}z^2 \frac{d}{dz} \nonumber \\
\hat{\sigma}_{\Lambda}(c) & = & -(\Lambda , \alpha_1 )^2-
(\Lambda ,\alpha_2 )^2-(\Lambda , \alpha_1 ) (\lambda , \alpha_2)
+\frac{2}{3}(\Lambda, \alpha_2-\alpha_1 )-\frac{1}{9} \nonumber
\end{eqnarray}
where $\alpha_1$ and $\alpha_2$ are the simple roots of $sl_3$
and $\Lambda$ is a weight of $sl_3$. It is easy to check that
these operators satisfy the algebra (\ref{alg}). For
$(\Lambda , \alpha_1 )= \frac{2}{3}(p-1)$ and $(\Lambda , \alpha_2 )
=\frac{2}{3}-\frac{1}{3}p-x$  the Fock representation
$\hat{\sigma}_{\Lambda}$ has a $p$ dimensional invariant subspace
\be
V=\{ P(z) \in {\bf C}[z] \mid (\frac{d}{dz})^pP(z)=0 \}
\ee
isomorphic to $W(p;x)$. This provides a Fock realization of the
representation $W(p;x)$.

In general (that is for arbitrary embeddings)
it is possible to find Fock realizations of finite $W$ algebras
by a generalized Miura transformation \cite{BGT}. Here we shall
not go into this however.

\section{Finite W symmetries in generalized Toda theories}
Consider the system of a particle moving on a group manifold $G$
($=SL(n)$). The action of such a particle can be taken to be
\be
S[g]=\frac{1}{2} \int dt \; \mbox{Tr}\left(
g^{-1}\frac{dg}{dt}g^{-1}\frac{dg}{dt}
 \right) \label{act}
\ee
where $g: {\bf R} \rightarrow G$ is the world line of the particle.
The equations of motion of this action are
\begin{equation}
\frac{d}{dt} \left( g^{-1}\frac{dg}{dt} \right) =  0, \label{eqm}
\end{equation}
or equivalently,
\begin{equation}
\frac{d}{dt} \left( \frac{dg}{dt}g^{-1} \right) =  0.
\end{equation}
In local coordinates $\{x^i\}$ the action looks like
\be
S=\frac{1}{2}\int dt \; g_{ij} \frac{dx^i}{dt}\frac{dx^j}{dt}
\ee
where $g_{ij}=R^a_iR^b_j K_{ab}$ and $g^{-1}\frac{dg}{dt}=R^a_i
\frac{dx^i}{dt}t_a$ (remember that $\{t_a\}$ is a basis of $g$
and $K_{ab}=Tr(t_at_b)$). From this we conclude that the action
(\ref{action}) discribes a free
particle moving in a curved background.

The action (\ref{action}) has a left and right $G$ symmetry
$g(t) \rightarrow a.g(t).b^{-1}$. From the equations of motion
we immediately find that
the conserved quantities are
\be
\bar{J}=\frac{dg}{dt}g^{-1}\equiv \bar{J}^at_a \mbox{  and  }J=
g^{-1}\frac{dg}{dt}  \equiv J^at_a
\ee
The conserved quantities $\{J^a\}$ form a Poisson algebra
\be
\{J^a,J^b\}=f^{ab}_cJ^c
\ee
(and similar equations for $\bar{J}$)
which, as we have seen, is nothing but the Kirillov Poisson
bracket written out in coordinates. Let's now consider what
happens to the theory (\ref{act}) when we reduce it.

Define $g^{(+)}= span\{t_{k,m}\}_{m>0},\;\; g^{(0)}=span\{t_{k,0}\},
\;\;g^{(-)}=span \{t_{k,m}\}_{m<0}$. Obviously $g=g^{(-)} \oplus
g^{(0)} \oplus g^{(+)}$. Let $G^{-},\; G^{0}, \; G^{+}$ be the
corresponding subgroups of $G$ and let $\pi_{\pm}$ be the
projections of $g$ onto $g^{(\pm )}$. The constraints we impose are (as
before)
\be
\pi_- (J) =t_- \;\;\; \mbox{and  } \pi_+ (\bar{J})=t_+  \label{one}
\ee
Inserting the generalized (local) Gauss decomposition
$g=g_-g_0g_+$ where $g \in G, \; g_{\pm} \in G^{\pm}$ and $g_0 \in G^0$
into eqns.(\ref{one}) we find
\begin{eqnarray}
g_0t_+g_0^{-1} & = & g_+^{-1}\frac{dg_+}{dt} \label{drie} \\
g_0^{-1}t_-g_0 & = & \frac{dg_-}{dt}g_-^{-1} \label{vier}
\end{eqnarray}
This means that the constrained currents look like
\begin{eqnarray}
J & = & g_-^{-1}(g_0^{-1}\frac{dg_0}{dt}+t_+)g_-
+g_-^{-1}\frac{dg_-}{dt}
\label{vijf} \\
\bar{J} & = & g_+(\frac{dg_0}{dt}g^{-1}_0+t_-)g_+^{-1}+\frac{dg_+}{dt}
g_+^{-1} \label{zes}
\end{eqnarray}
Note that now the equations of motion (\ref{eqm}) can be written
as
\be
[\frac{d}{dt},\frac{d}{dt}+J]=0
\ee
Conjugating this equation by $g_-$ , using eqns. (\ref{drie}),
(\ref{vier}), (\ref{vijf}) and writing out the commutator we find
\be
\frac{d}{dt}\left( g_0^{-1}\frac{dg_0}{dt} \right)
=[g_0^{-1}t_-g_0,t_+] \label{star}
\ee
This evolution equation describes the gauge invariant part of the
constrained theory. The action corresponding to this equation is
\be
S[g_0]=\frac{1}{2}\int dt\; \mbox{Tr}\left( g_0^{-1}\frac{dg_0}{dt}
g_{0}^{-1}\frac{dg_0}{dt} \right) + \int dt \; \mbox{Tr}\left(
g_0^{-1}t_-g_0t_+ \right) \label{action}
\ee
which describes a particle moving on the subgroup $G_0$ of $G$
with some selfinteraction. It can be shown that the theory (\ref{action})
has the nonlinear 'finite' $W$ symmetry corresponding to the  $sl_2$
subalgebra $\{t_3,t_{\pm}\}$ of $g$.

Stictly speaking the above arguments only work when the $sl_2$
subalgebra which one considers provides an 'integral grading'
of the Lie algebra $g$ because then there are only first class
constraints. However it can be shown that it is always possible
to find a set of first class constraints that give the same
constrained and gauge fixed manifold $g_{fix}$. This is done
by imposing
only 'half' (there are always an even number of second
class constraints) of the constraints
that turned out to be second class such that they become first
class. The other half can then be imposed as gauge fixing conditions
(see for a treatment of this in the present context \cite{D4}).

In the case where the $sl_2$ embedding is the 'principal' embedding
of $sl_2$ into $g=sl_n$ \cite{Dynkin} equation (\ref{star}) reduces to
the ordinary finite Toda equations
\be
\frac{d^2q_i}{dt^2}+\mbox{exp}\left( \sum_{j=1}^{n-1}K_{ij}q_j \right)
=0
\ee
where $i=1, \ldots , rank(g)=n-1$,  $K_{ij}$ is the Cartan
matrix of $sl_n$ and $g_0=exp(q_iH_i)$..

The generalized finite Toda theories (\ref{star}) were already derived
in \cite{BV} as dimensional reductions of the selfdual Yang-Mills
equations. For some examples the solutions were constructed, however
the general solution space is to our knowledge still not known. The finite
$W$ algebras may provide a new tool in this research since they
are expected to transform solutions of (\ref{star}) to (different)
solutions of (\ref{star}) just like the general solution of (\ref{eqm})
can be found by letting the symmetry group $G \times G$ act on the
simplest solution $g=e^{tX}$.

\section*{Acknowledgements}
I would like to thank Sander Bais, Jan de Boer,  Jacob Goeree
and Klaas Landsman
for many conversations on the subjects presented in this
paper.

\end{document}